# Magnetic vortices as efficient nano heaters in magnetic nanoparticle hyperthermia.

N. A. Usov[1,2,3], M. S. Nesmeyanov[3] and V. P. Tarasov[1]

[1]*National University of Science and Technology «MISiS», 119049, Moscow, Russia*
[2]*Pushkov Institute of Terrestrial Magnetism, Ionosphere and Radio Wave Propagation, Russian Academy of Sciences, IZMIRAN, 142190, Troitsk, Moscow, Russia*
[3]*National Research Nuclear University "MEPhI", 115409, Moscow, Russia*

**Abstract**

Magnetic vortices existing in soft magnetic nanoparticles with sizes larger than the single-domain diameter can be efficient nano-heaters in biomedical applications. Using micromagnetic numerical simulation we prove that in the optimal range of particle diameters the magnetization reversal of the vortices in spherical iron and magnetite nanoparticles is possible for moderate amplitudes of external alternating magnetic field, $H_0 < 100$ Oe. In contrast to the case of superparamagnetic nanoparticles, for the vortex configuration the hysteresis loop area increases as a function of frequency. So that high values of the specific absorption rate, on the order of 1000 W/g, can be obtained at frequencies $f = 0.5 - 1.0$ MHz. Because the diameter $D$ of a non single-domain particle is several times larger than the diameter $d$ of a superparamagnetic particle, the volume of heat generation for the vortex turns out to be $(D/d)^3$ times larger. This shows the advantage of vortex configurations for heat generation in alternating magnetic field in biomedical applications.

## I. INTRODUCTION

The performance of magnetic nanoparticles to generate heat in alternating external magnetic field can be used in magnetic nanoparticle hyperthermia [1-6] for local heating of biological tissues to suppress the growth and destroy tumors. Magnetic nanoparticles can also be used in targeted drug delivery, as release of the drug in the designated place from the corresponding insulating envelope can be achieved [7] by remote destruction of the envelope by heating. It is desirable to get useful therapeutic effect with as low as possible concentration of magnetic nanoparticles in biological media. Therefore, only nanoparticles with sufficiently large value of the specific absorption rate (SAR) are suitable for magnetic nanoparticle hyperthermia.

For biomedical applications it is important also to ensure the magnetization reversal of an assembly of magnetic nanoparticles in an alternating magnetic field of moderate amplitude, $H_0 < 100 - 200$ Oe. Indeed, the use of strong alternating magnetic field requires generation of sufficiently large electric currents. It can be dangerous in a clinic. In addition, according to the empirical Brezovich's criterion [6,4], the alternating magnetic field is harmless to the human body if its amplitude and frequency $f$ satisfy the condition $fH_0 < 5\times10^9$ A/(ms). Therefore, in recent experimental and theoretical studies in magnetic hyperthermia [7-20] the behavior in alternating magnetic field of assemblies of superparamagnetic nanoparticles with diameters substantially smaller than the single-domain diameter $D_c$ is mostly investigated, as at a room temperature the coercive force of superparamagnetic nanoparticles with diameters $d < D_c$ decreases significantly under the influence of thermal fluctuations of their magnetic moments.

In this paper, we draw attention to a possibility of using non single-domain nanoparticles with diameters $D > D_c$ for local heating of the biological media. It is well known [21-27] that for nanoparticles of soft magnetic type vortex configuration has the lowest total energy for diameters $D > D_c$. Of course, the average magnetization of the vortex rapidly decreases with increasing particle diameter. However, it remains appreciable, $<M>/M_s > 0.3 - 0.5$, for particles with diameters close to the single-domain diameter.

At present, the nanoparticles of iron and iron oxides are considered to be most promising for use in magnetic hyperthermia mainly because of their low toxicity [28]. Therefore, we study theoretically the behavior of vortices in spherical iron and magnetite nanoparticles in alternating magnetic field. In the case of magnetite, the nanoparticles of cubic shape are also considered. In the experiment, sufficiently large cubic nanoparticles can be obtained by various methods [12,14,16,29-31]. Apparently, these particles have a perfect crystal structure [30], as their magnetic characteristics are close to the corresponding values for bulk material [32].

It has been found recently [5,33-35] that being embedded in a biological environment, for example, into a tumor, magnetic nanoparticles turn out to be tightly bound to the surrounding tissues. Therefore, the rotation of magnetic nanoparticles as a whole under the influence of alternating external magnetic field is greatly hindered. In such a case, the Brownian relaxation is unimportant [5], and only the evolution of the particle magnetization under the influence of an alternating magnetic field has to be considered. In this paper the numerical simulations of vortex configurations and low frequency hysteresis loops are carried out based on the solution of the Landau-Lifshitz-Gilbert (LLG) equation [25,36]. It is shown that in an optimal range of particle diameters the magnetization reversal of the vortex is possible for moderate amplitudes of external alternating magnetic field, $H_0 < 100 - 200$ Oe. In contrast to the case of superparamagnetic nanoparticles [2,5,37], the area of the low frequency hysteresis loops for vortex increases with increase of alternating field frequency. As a result, very large SAR values, of the order of 1000 W/g, have been obtained at frequencies $f \sim 0.5 - 1.0$ MHz. It is worth mentioning also that because the diameter $D$ of a non single-domain particle is several times larger than the diameter $d$ of a superparamagnetic particle, for vortex configuration the volume of heat generation turns out to be $(D/d)^3$ times larger.

These results seem important for further successful development of magnetic nanoparticle hyperthermia.

## II. Numerical simulation

Dynamics of the unit magnetization vector $\vec{\alpha}(\vec{r})$ of a non single domain nanoparticle in applied magnetic field $\vec{H}_0 \sin(\omega t)$ is described by the LLG equation [25,36]

$$\frac{\partial \vec{\alpha}}{\partial t} = -\gamma \left( \vec{\alpha} \times \vec{H}_{ef} \right) + \kappa \left( \vec{\alpha} \times \frac{\partial \vec{\alpha}}{\partial t} \right), \quad (1)$$

where $\gamma$ is the gyromagnetic ratio and $\kappa$ is the phenomenological damping constant. The effective magnetic field $\vec{H}_{ef}$ acting on the unit magnetization vector can be calculated as a derivative of the total nanoparticle energy [36]

$$\vec{H}_{ef} = -\frac{\partial W}{V M_s \partial \vec{\alpha}},$$

$$M_s \vec{H}_{ef} = C \Delta \vec{\alpha} - \frac{\partial w_a}{\partial \vec{\alpha}} + M_s \left( \vec{H}_0 \sin(\omega t) + \vec{H}' \right). \quad (2)$$

Here $V$ is the nanoparticle volume, $M_s$ is the saturation magnetization, $C$ is the exchange constant, and $\vec{H}'$ is the demagnetizing field. The magneto-crystalline anisotropy energy density of a nanoparticle with cubic anisotropy is given by [36]

$$w_a = K_c \left( \alpha_x^2 \alpha_y^2 + \alpha_x^2 \alpha_z^2 + \alpha_y^2 \alpha_z^2 \right), \quad (3)$$

where $K_c$ is the cubic anisotropy constant.

For numerical simulation a non single-domain nanoparticle is approximated by a set of small ferromagnetic cubes of side $b$ much smaller than the exchange length $L_{ex} = \sqrt{C}/M_s$ of the ferromagnetic material. Typically, several thousands of numerical cells, $N \sim 10^3 - 10^4$, is necessary to approximate with sufficient accuracy the vortex type magnetization distribution in nanoparticle volume. For reliable calculation of the low frequency hysteresis loops of the nanoparticle it is important to keep the numerical time step $\Delta t$ sufficiently small [37] with respect to the characteristic precession time of the unit magnetization vectors in various numerical cells, $T_p \sim 1/\gamma <H_{ef}>$, where $<H_{ef}>$ is the average value of the effective magnetic field of the nanoparticle. In the present calculations the numerical time step is fixed at $\Delta t/T_p = 1/30$, the magnetic damping parameter being $\kappa = 0.5$.

The equilibrium micromagnetic configurations for the nanoparticles studied were calculated using the same LLG equation with zero applied magnetic field. In accordance with the Eq. (1), the final magnetization state is assumed to be stable under the condition

$$\max_{(1 \leq i \leq N)} \left\| \vec{\alpha}_i \times \vec{H}_{ef,i} / \|\vec{H}_{ef,i}\| \right\| < 10^{-6}, \quad (4)$$

where $\vec{\alpha}_i$ and $\vec{H}_{ef,i}$ are the unit magnetization vector and effective magnetic field in the $i$-th numerical cell, respectively.

## III. Results and discussion

**Spherical iron nanoparticles**

Let us first consider iron nanoparticles, which are especially interesting for use in magnetic hyperthermia [9-11] due to the high saturation magnetization, $M_s = 1700$ emu/cm$^3$. The cubic magnetic anisotropy constant of iron is assumed to be [32] $K_c = 4.6 \times 10^5$ erg/cm$^3$, the exchange constant is $A = C/2 = 2.0 \times 10^{-6}$ erg/cm. Fig. 1 shows the energy diagram of stable micromagnetic states existing in a spherical iron nanoparticle with cubic anisotropy, depending on its diameter. As Fig. 1 shows, for iron nanoparticles the single-domain diameter is approximately $D_c \approx 26$ nm. This value, determined numerically, is in agreement with the analytical estimates [38]. The insert in Fig. 1 shows that in the range of diameters $D = 30-40$ nm the average reduced magnetization of the vortex, although decreasing as a function of the particle diameter, remains sufficiently large, $<M>/M_s > 0.3$.

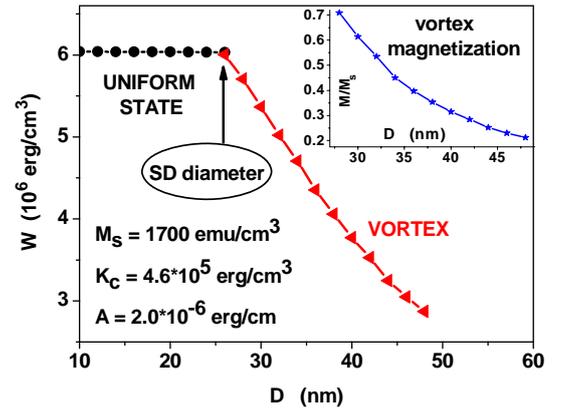

Fig. 1. Energy diagram of equilibrium magnetization distributions in a spherical iron nanoparticle with cubic anisotropy. The insert shows the reduced average magnetic moment of the vortex as a function of the nanoparticle diameter.

The magnetization distribution for the vortex in iron nanoparticle of diameter $D = 42$ nm, obtained by means of numerical simulation, is shown in Fig. 2. In the cylindrical core of the vortex the particle magnetization remains approximately homogeneous. This region gives the main contribution to the remanent vortex magnetization. The outer shell of the vortex where the magnetization has an azimuthal direction does not contribute to the remanent magnetization of the particle.

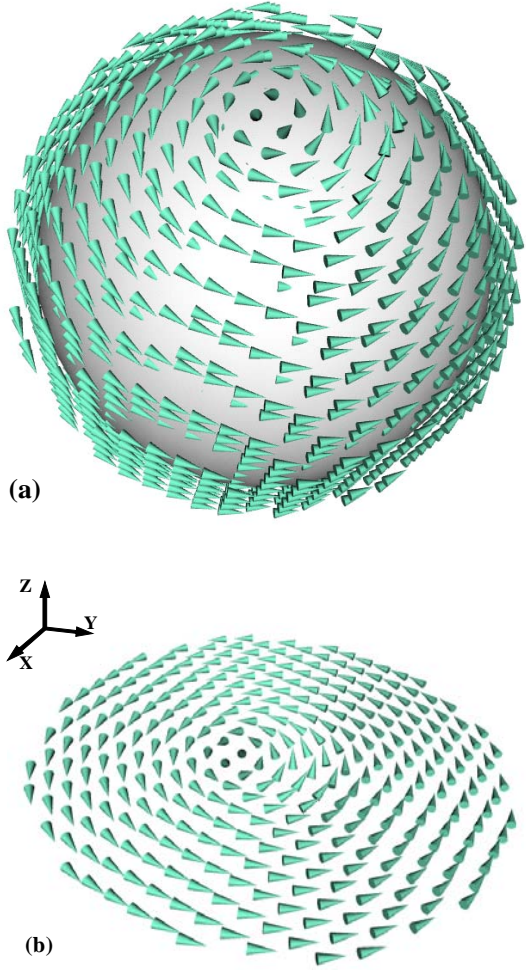

Fig. 2. a) General view of the vortex in iron nanoparticle of diameter $D = 42$ nm; b) the diametrical cross section of the vortex perpendicular to the easy anisotropy axis.

It is well known [36], that for a uniformly magnetized iron nanoparticle with positive cubic anisotropy constant, $K_c > 0$, the easy anisotropy axes are parallel to the Cartesian coordinate axes, i.e. (±1,0,0), etc. It is interesting to note that in a small range of diameters, $D_c < D < 32$ nm, the vortex axis is oriented along one of the cube diagonals. However, as the diameter of the nanoparticle increases, $D > 32$ nm, the total energy of the vortex decreases if its axis is directed along one of the easy axes of cubic anisotropy.

To calculate the low-frequency hysteresis loops of iron nanoparticles with diameters $D > D_c$ it is necessary to study the vortex dynamics in alternating magnetic field. Note that for particles with cubic anisotropy, the region of nonequivalent directions of the external magnetic field with respect to the directions of the easy anisotropy axes of the spherical coordinates ($\psi$, $\omega$) is bounded [25] by the spherical triangle $\Omega$: $0 \leq \psi_h \leq \pi/4$; $0 \leq \omega_h \leq \arctan(1/\cos(\psi_h))$. Therefore, it is sufficient to study the low-frequency hysteresis loops of vortices for directions of the external magnetic field within this spherical triangle.

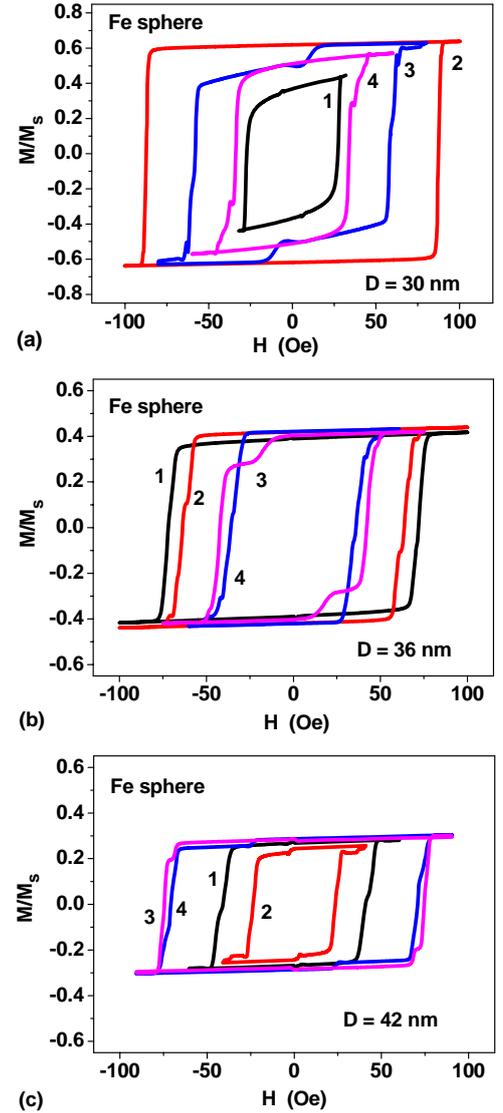

Fig. 3. Hysteresis loops of vortex in iron nanoparticles of different diameters for some characteristic directions of applied magnetic field: 1) $\omega_h = \psi_h = 0.0$; 2) $\omega_h = 0.955$, $\psi_h = \pi/4$; 3) $\omega_h = \pi/4$, $\psi_h = 0.0$; 4) $\omega_h = \psi_h = \pi/8$. The frequency and amplitude of alternating magnetic field are given by $f = 1$ MHz and $H_0 = 100$ Oe, respectively.

Fig. 3 shows calculated hysteresis loops of iron nanoparticles of different diameters, $D = 30$, 36 and 42 nm, respectively. The loops in Fig. 3 are calculated for some characteristic directions of the magnetic field lying in the corners and inside the spherical triangle $\Omega$. One can see that in all particular cases considered the particle coercive force at the frequency $f = 1$ MHz does not exceed 100 Oe, which is appealing for application in magnetic hyperthermia. As Fig. 3 shows, with increasing nanoparticle diameter, the hysteresis loop area gradually decreases, as the average magnetization of the vortex decreases as a function of particle diameter. Nevertheless, it remains large enough even for nanoparticles of diameter $D = 42$ nm. As shown below (see Fig. 7), in the range of diameters $D = 30 - 42$ nm, the calculated SAR of a dilute random assembly of iron nanoparticles is of the order of 1000 W/g at frequencies $f > 0.5$-1 MHz.

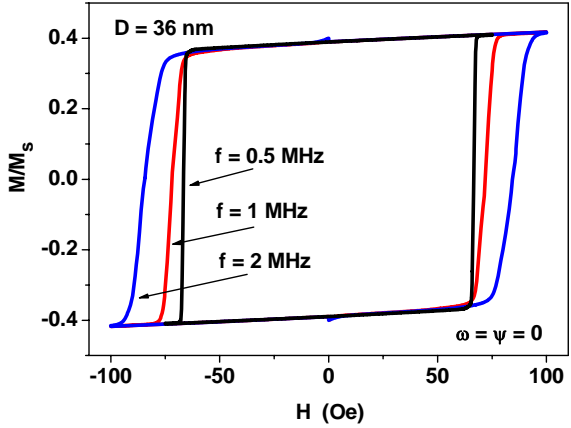

Fig. 4. The hysteresis loops of iron nanoparticle with diameter $D = 36$ nm for various frequencies at $\omega_h = \psi_h = 0.0$.

It is interesting to note that, in contrast to superparamagnetic nanoparticles, for which the area of the hysteresis loop at moderate values of the alternating field amplitude, $H_0 \leq 100$ Oe, falls [37] with increasing frequency in the interval $f > 0.5 - 1$ MHz, the vortex hysteresis loop area grows as a function of the frequency. As an example, Fig. 4 shows the behavior of the vortex hysteresis loops at a fixed amplitude $H_0 = 100$ Oe and various frequencies for iron nanoparticle with diameter $D = 36$ nm for one of the characteristic directions of the alternating magnetic field in the spherical triangle $\Omega$. Similar results were also obtained for particles of different diameters and different directions of the alternating magnetic field. Therefore, for vortex it is reasonable to increase the frequency of the alternating magnetic field, while decreasing simultaneously the field amplitude. Note that decreasing of the alternating field amplitude is preferable for technical reasons, since this reduces the cost and increases the safety of the equipment that will be used in medical practice.

**Spherical magnetite nanoparticles**

Similar results were obtained for spherical nanoparticles of magnetite, $Fe_3O_4$, which are considered [1-5] to be the most suitable for biomedical applications. The saturation magnetization of magnetite nanoparticles was assumed equal to the bulk value [32], $M_s = 480$ emu/cm$^3$, the cubic magnetic anisotropy constant is negative, $K_c = -1.0 \times 10^5$ erg/cm$^3$, the exchange constant is $A = 1.0 \times 10^{-6}$ erg/cm. For particles with negative cubic anisotropy constant the easy anisotropy axes are parallel to the diagonals of the cube. As Fig. 5 shows, in agreement with analytical estimates [38] the single-domain diameter of magnetite nanoparticle, $D_c \approx 64$ nm, is much higher as compared to that of iron nanoparticle because of significantly lower saturation magnetization of magnetite.

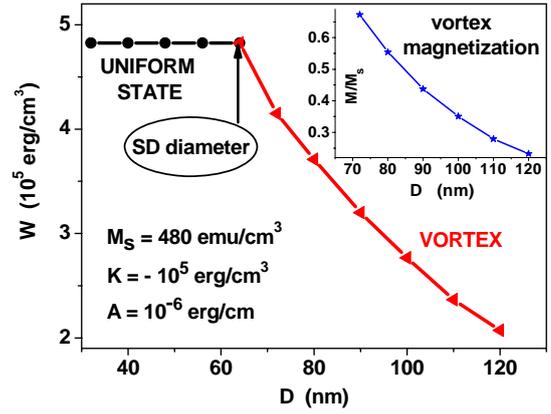

Fig. 5. Energy diagram of equilibrium magnetization states in a spherical magnetite nanoparticle. The insert shows the average reduced magnetic moment of the vortex in magnetite nanoparticle depending on the particle diameter.

As inset in Fig. 5 shows, in the range of diameters $D = 70 - 100$ nm, the average magnetization of magnetite nanoparticles remains appreciable. This should result in low-frequency hysteresis loops of a sufficiently large area. As an example of calculations performed, Fig. 6 shows hysteresis loops of magnetite nanoparticles with diameters $D = 72$ nm and 80 nm at frequency $f = 1$ MHz and $H_0 = 100$ Oe for some characteristic directions of the alternating magnetic field in the spherical triangle $\Omega$.

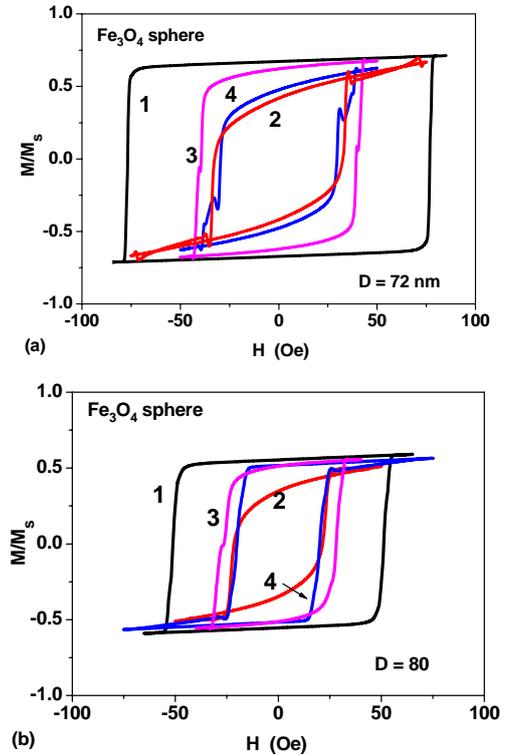

Fig. 6. The vortex hysteresis loops for spherical magnetite nanoparticles with diameters $D = 72$ nm (a) and $D = 80$ nm (b) for some characteristic directions of the external magnetic field: 1) $\omega_h = \psi_h = 0.0$; 2) $\omega_h = 0.955$, $\psi_h = \pi/4$; 3) $\omega_h = \pi/4$, $\psi_h = 0.0$; 4) $\omega_h = \psi_h = \pi/8$.

One can see in Fig. 6, that the coercive force of the calculated hysteresis loops does not exceed 100 Oe. Similar results were obtained for magnetite nanoparticles in the range of diameters $D$ = 70-100 nm.

For each particular hysteresis loop its area $A$ has been calculated in the variables ($M/M_s$, $H$). The corresponding SAR is then calculated [2,5] as $SAR = 10^{-7} M_s f A / \rho$ (W/g), where $\rho$ is the nanoparticle density. The latter is given by $\rho$ = 7.88 g/cm$^3$ for iron nanoparticles and $\rho$ = 5.0 g/cm$^3$ for magnetite nanoparticles, respectively.

Having calculated a sufficient number of the low frequency hysteresis loops with the direction of the magnetic field within the spherical triangle $\Omega$, it is possible to calculate the SAR of a dilute randomly oriented assembly of magnetic nanoparticles being in the vortex states. Fig. 7 shows the dependence of SAR of dilute randomly oriented assemblies of spherical iron and magnetite nanoparticles, as well as cubic magnetite nanoparticles as a function of particle size. The SAR averaging was carried out over 10 - 12 independent magnetic field directions within the spherical triangle $\Omega$. As shown in Fig. 7, for a dilute assembly of iron nanoparticles in the range of diameters $D$ = 30 - 42 nm the SAR exceeds the value of 1000 W/g at a frequency $f$ = 1 MHz. For magnetite nanoparticles, the SAR values also remain high, at least for nanoparticle diameters close to the single-domain one.

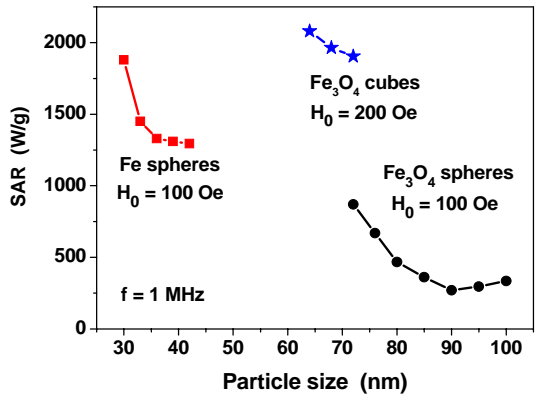

Fig. 7. SAR of dilute randomly oriented assemblies of iron and magnetite nanoparticles in vortex states depending on the particle diameter.

Note that the use of non-single-domain magnetic nanoparticles can be advantageous in magnetic hyperthermia also because the physical volume of heat generation for such nanoparticles is significantly greater than that for the corresponding superparamagnetic nanoparticles. For example, as shown in Ref. 37, for slightly elongated superparamagnetic magnetite nanoparticles, the maximum SAR at frequencies of the order of $f$ = 0.5 MHz occurs for particles with diameter $d$ = 13 nm. On the other hand, for vortex in magnetite nanoparticles the diameters $D$ = 70 - 75 nm are optimal. Accordingly, the ratio of the volumes of these particles is $(D/d)^3 \approx 160$. As Fig. 8 shows, a large number of small superparamagnetic nanoparticles of diameter $d$ can be distributed in the volume of a sphere of diameter $D$. However, the experiment and calculations [5,11,39-41] show that the magneto-dipole interaction of closely located superparamagnetic nanoparticles significantly decreases the SAR value for such a cluster of nanoparticles.

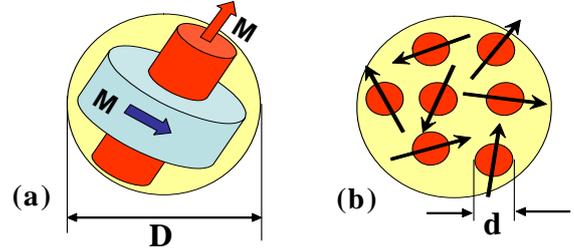

Fig. 8. The volumes for heat generation for a vortex in a nanoparticle of diameter $D > D_c$ (a), and for a cluster of superparamagnetic nanoparticles (b) with significantly smaller diameters $d < D_c$.

**Magnetite nanocubes**

Modern methods of chemical synthesis of magnetic particles allow the growth of magnetite nanoparticles of cubic shape [12,14,16,29,30] with the edge size $L$ from 18 to 160 nm. Moreover, these magnetite nanoparticles are of perfect quality [30] as their saturation magnetization is close to the corresponding bulk value [32]. Therefore, we carried out also the calculations of the equilibrium magnetization distributions and low frequency hysteresis loops for magnetite nanoparticles of cubic shape.

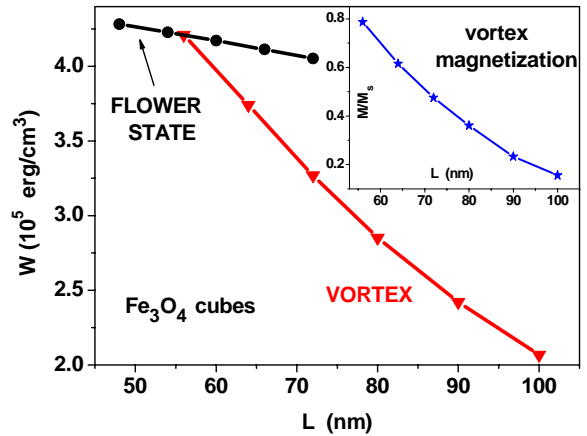

Fig. 9. Energy diagram of the equilibrium magnetization states in a cubic magnetite nanoparticle. The insert shows the average reduced magnetic moment of the vortex depending on the length of the cube's side.

As Fig. 9 shows, quasi-uniform flower state [42,43] exists in sufficiently small cubic nanoparticles of soft magnetic type. At larger sizes, the vortex competes in energy with the flower state. The intersection of the curves corresponding to the flower state and vortex determines the effective single-domain size [25] for cubic-shaped nanoparticles. According to Fig. 9, for cubic magnetite nanoparticle the effective single-domain size is $D_{c,ef}$ = 56 nm. It is smaller than the single-domain diameter of spherical magnetite nanoparticle, $D_c$ = 64 nm.

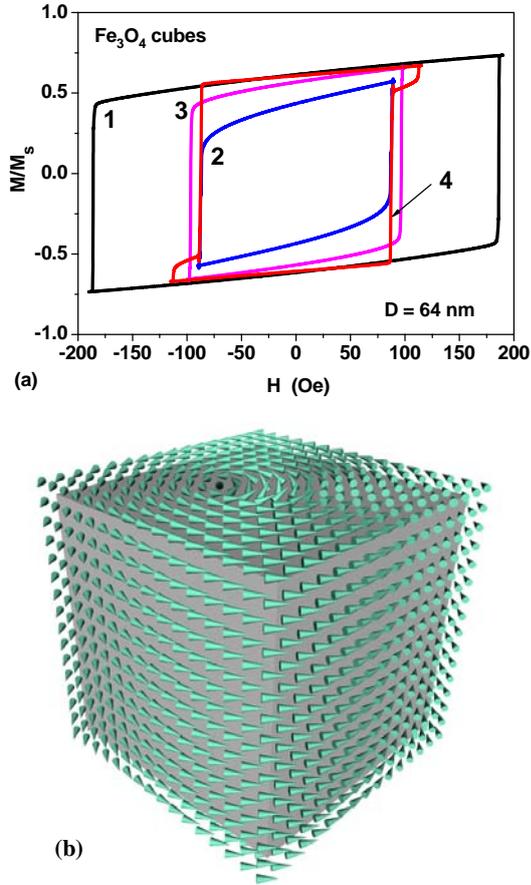

Fig. 10. a) Hysteresis loops of vortex configuration in a cubic magnetite nanoparticle with size $L$ = 64 nm for some characteristic directions of external magnetic field: 1) $\omega_h = \psi_h = 0.0$; 2) $\omega_h = 0.955$, $\psi_h = \pi/4$; 3) $\omega_h = \pi/4$, $\psi_h = 0.0$; 4) $\omega_h = \psi_h = \pi/8$. b) the vortex magnetization distribution in a magnetite nanoparticle with size $L$ = 64 nm.

Fig. 10a shows the hysteresis loops of a cubic magnetite nanoparticle with the size $L$ = 64 nm for some characteristic directions of an external magnetic field in a spherical triangle Ω. The frequency and amplitude of the alternating magnetic field are given by $f$ = 1 MHz and $H_0$ = 200 Oe, respectively. Fig. 10b shows an example of the vortex magnetization distribution in cubic magnetite nanoparticle in the ground state, in the absence of an external magnetic field.

In the case of cubic magnetite nanoparticles the coercive force of low frequency hysteresis loops for certain directions of the external magnetic field turns out to be larger than that for the case of spherical nanoparticles. To obtain complete hysteresis loops it was necessary to increase the amplitude of the alternating magnetic field to a value $H_0$ = 200 Oe. Fig. 7 also shows the SAR of dilute randomly oriented assembly of cubic magnetite nanoparticles as a function of the cube side.

As can be seen from Fig. 7, in the investigated range of sizes $L$ = 64 - 72 nm the SAR of the assembly reaches values of the order of 2000 W/g at the frequency $f$ = 1 MHz and amplitude $H_0$ = 200 Oe.

## IV. Conclusions

The vortex magnetization distributions in magnetically soft nanoparticles have long been studied both theoretically and experimentally [21-27]. These states are of fundamental importance, as the vortices have the lowest total energy in these nanoparticles of sufficiently large diameters. However, up to now, vortices in magnetically soft nanoparticles have not found practical application. It seems therefore interesting if the vortices can be used to generate heat in an alternating magnetic field in magnetic hyperthermia.

In this paper the low-frequency hysteresis loops of vortices existing in iron and magnetite nanoparticles have been investigated by means of numerical simulation. It is shown that for a dilute assemblies of these nanoparticles in the optimal range of diameters it is possible to obtain very high SAR values, on the order of 1000 W/g at frequencies $f$ ~ 0.5-1.0 MHz and moderate alternating magnetic field amplitudes, $H_0$ < 100 Oe. For technical and medical requirements it is desirable to reduce the amplitude of the alternating magnetic field acting on the nanoparticles. To maintain sufficiently high SAR values, the decrease in the alternating field amplitude can be compensated by increasing field frequency. In this connection, the use of vortices in magnetic hyperthermia may be preferable, since with increasing frequency the hysteresis loop area for vortices increases, in contrast to the assembly of superparamagnetic nanoparticles. In the latter case for magnetic field amplitudes, $H_0$ < 100 Oe, the loop area decreases [37] with increasing frequency in the interval $f$ > 0.5 - 1.0 MHz.

It is also important to note that in the case of non single-domain nanoparticles with diameters $D > D_c$ the volume of the heat generation is $(D/d)^3$ times larger than that for small superparamagnetic nanoparticle of diameter $d < D_c$. In a dense assembly of superparamagnetic nanoparticles a significant decrease of the SAR occurs [39-41] under the influence of mutual magneto-dipole interactions of the nanoparticles. At the same time, one can expect that the effect of the magnetostatic interactions will not be so significant for the vortex states, since the magnetic moments of the vortices are relatively small.

## Acknowledgments

The authors gratefully acknowledge the financial support of the Ministry of Education and Science of the Russian Federation in the framework of Increase Competitiveness Program of NUST «MISIS», contract № K2-2015-018.